\documentclass[twocolumn,preprintnumbers,amsmath,amssymb]{revtex4}

\usepackage{graphicx}
\usepackage{dcolumn}
\usepackage{bm}

\begin{document}


\title{On the connection between structural distortion and magnetism in
graphene with a single vacancy}

\author{Wendel S Paz$^1$, Wanderl\~a L Scopel$^{1,2}$, Jair C C
Freitas$^1$}
\affiliation{$^1$Departamento de F\'{\i}sica, Universidade Federal do Esp\'
irito Santo, Vit\' oria, Brazil }
\affiliation {$^2$Departamento de Ci\^encias Exatas, Universidade Federal
Fluminense,
Volta Redonda, RJ, Brazil}
\email{wlscopel@gmail.com}

\begin{abstract}

 The correlation between structural distortion and emergence of magnetism in
graphene containing a single vacancy was investigated using first-principles
calculations based on density functional theory (DFT). Our results
have shown that a local distortion is formed around the vacancy, with
reconstruction of two atomic bonds and with a dangling bond remaining at the
third atom adjacent to the vacancy. A systematic investigation of the possible
out-of-plane displacement of this third atom was then carried out, in order to
ascertain its effects on the magnetic features of the system. The ground state
was definitely found to be magnetic and planar, with spin-resolved $\sigma$ and
$\pi$ bands contributing to the total magnetic moment. However, we have also
found that metastable solutions can be achieved if an initial shift of the third
atom above a minimum threshold from the graphene plane is provided, which leads
to a non-planar geometry and a non-magnetic state.
 \end{abstract}
 

\maketitle

\section{Introduction}

The issue of magnetism in graphene and related materials-including graphene
multilayers, graphite,nanocarbons and others-has been the focus of intense
research in the last years, both from the experimental and the theoretical
points of view \cite{helm,yazyev2,oeiras,esquinaz,monbru}. The onset of
intrinsic magnetic properties is always linked to some kind of defects
introduced in the bipartite lattice, such as atomic vacancies, chemisorbed
species or edge sites \cite{yazyev2,oeiras, cervenka}. Single vacancies
cause the appearance of dangling bonds and, depending on the defect
concentration and on the degree of passivation associated with eventually
present hydrogen atoms, the magnetic moments due to electrons with
uncompensated spins in $\sigma$ and $\pi$ orbitals can couple to give a total
magnetic moment in the range 1-2 $\mu_B$ per atomic vacancy \cite{helm, yazyev2,
palacios}. Although there have been some experimental reports suggesting the
existence of ferromagnetism at room temperature in defective graphene
\cite{wang, ugeda}, recent work using measurements of magnetic
properties \cite{nair} and muon spin rotation spectroscopy \cite{rico} has ruled
out this possibility. In particular, Nair et al. \cite{nair} showed that
localized defects carry magnetic moments with spin $\frac{1}{2}$, leading to
strong paramagnetism but no magnetic order down to liquid He temperature. Some
recent calculations performed by Palacios $\&$ Yndur\'ain for variable defect
concentrations in graphene sheets containing single vacancies have confirmed
these findings, by showing that the vacancy-induced magnetism associated with
extended $\pi$ orbitals has a trend to vanish at any experimentally relevant
vacancy concentration-thus spoiling any possibility of ferromagnetic or
antiferromagnetic interactions associated with those $\pi$ electrons
\cite{palacios}. On the other hand, this work also showed that localized,
non-interacting states associated with a flat $\sigma$ band are the only ones
surviving for defect concentrations approaching zero, in apparent agreement with
the paramagnetic properties experimentally observed for graphene irradiated with
protons \cite{nair}.

When a single atomic vacancy is formed in an otherwise perfect two-dimensional
graphene lattice in the absence of hydrogen (or other gases that could lead to
passivation), three dangling bonds remain in the atoms surrounding the vacancy. 
A Jahn-Teller distortion in this neighbourhood leads to the reconstruction
of the bond between two of the atoms closest to the vacancy, leaving the third
atom less bound to the network and with an unsaturated dangling bond
\cite{helm,faccio}, as illustrated in Figure \ref{structure}. The question then
arises whether or not this atom (identified as atom 3 in Figure 1) will remain
in the plane, an issue that has received a number of apparently conflicting
responses in the last years. El-Barbary et al. used the local density
approximation (LDA) in an atomic cluster to model a defective
hydrogen-terminated graphene ribbon \cite{barbary}. They found that
the fully optimized structure is distorted, with the atom with the unsaturated
dangling bond moving out of the plane by 0.47 \AA. When using calculations with
spin polarization, the spin-polarized state was found to be higher in 
energy than the unpolarized one by ca. 0.5 eV, leading the authors to
conclude that the ground state of vacancy-containing graphite should be
nonmagnetic \cite{barbary}. A similar method was followed by Dharma-wardana $\&$
Zgierski, who used DFT calculations to study the structural, electronic and
magnetic properties of a single vacancy in a finite graphene fragment with
zig-zag edges terminated with hydrogen atoms \cite{wardana}. They
found the ground state was spin polarized and planar, lying  ca. 0.2 eV below
the non-planar unpolarized structure, with the latter showing a 
saddle-like distortion around the vacancy. In contrast, when using
periodic boundary conditions to study an infinite graphene sheet containing a
single vacancy, no out-of-plane distortions were found in the fully relaxed
structure, no matter whether or not spin polarization was used in the
calculations \cite{wardana}. In a related work, Ma et al., employing the
generalized gradient approximation (GGA), concluded that the ground state
of graphene containing a single vacancy was spin polarized, with the atom with
no reconstructed bonding displaced by 0.18 \AA out of the plane \cite{ma}.
When forcing a nonmagnetic solution, which was found to be higher in energy by
0.1 eV as compared to the ground state, they found this displacement being
increased to 0.46 \AA, in agreement with the results of El-Barbary et al.
\cite{barbary}. Accordingly, Dai et al. also found, using the same GGA method, a
magnetic moment of 1.33 $\mu$ B for a monovacancy in graphene and reported an
upward shift of atom 3 from the plane by 0.184 \AA \cite{dai}. 
These results are in contrast with the very recent GGA calculations
reported by Nanda et al., who found no displacement of any atom out of the
graphene plane containing a single vacancy \cite{nanda}.

A more detailed investigation of this issue was reported by Faccio et al.,
who used GGA calculations to compare the effects caused by carbon vacancies and
boron doping in graphene\cite{faccio2}. By studying monovacancies in
graphene sheets of different sizes (and thus with variable degree of
interaction between vacancies in different supercells), these authors found
that a nonmagnetic solution was higher in energy by ca. 0.4 eV as compared to
the magnetic structure for the whole range of defect concentrations. Also, the
establishment of the nonmagnetic solution was accompanied by an out-of-plane
displacement of atom 3 by 0.3 \AA. Interestingly, it was reported that when this
nonmagnetic solution was used as the starting point of a new spin-polarized
calculation, allowing full structural relaxation, then the structure relaxed
back to the flat graphene sheet, with a net magnetic moment in the range 1.0-1.3
$\mu_B$, depending on the defect concentration \cite{faccio2}. 
The results recently reported by Ugeda et al. in an investigation about the
effects of single vacancies in a graphene monolayer deposited on Pt surface
\cite{ugeda2} also helped to shed some light into this subject. In order to
understand the absence of magnetism in the graphene/Pt system, the authors 
first studied the features of a pure graphene sheet containing a single vacancy;
again, spin-polarized calculations showed that the ground state was magnetic
(with a magnetic moment of 1.5 $\mu_B$) and nearly planar, with a nonmagnetic
solution found 0.1 eV above the ground state. However, by systematically forcing
an out-of-plane displacement of atom 3, a decrease in the magnetic moment as a
function of the displacement was obtained, with a nonmagnetic state being found
(even using spin-polarized calculations) for a displacement of ca. 0.5 \AA. This
quenching of magnetic moment was attributed to the mixing of $\sigma$ and $\pi$
states in the spin-resolved density of states (DOS), causing a partial or total
compensation of the magnetic moments associated with the former $\sigma$
and $\pi$ orbitals \cite{ugeda2}. This conclusion was in agreement with
the work by Dharma-wardana $\&$ Zgierski, who reported that in the unpolarized 
structure of a finite graphene fragment with a single vacancy the distortion
caused a disconnection of the electron associated with the vacancy from
the remaining 2D electron network; this led to the formation of two singlet
electron pairs - one associated with the two electrons at atom 3 and the 
other one associated with the reconstructed bonding between atoms 1 and 2
(see Fig. \ref{structure}), causing the structure to become nonmagnetic
\cite{wardana}.

Considering this scenario with some apparently discrepant conclusions as well as
a number of sparse complementary findings, we decided to
undertake a systematic investigation of the effects of out-of-plane 
displacements on the structural, electronic and magnetic features of a graphene
sheet containing a single vacancy. Both spin-polarized and unpolarized
calculations with full structural relaxation were performed, allowing us to
compare the total energies, magnetic moments and electronic configurations of
each relaxed structure corresponding to different initial displacements. These
results unequivocally show that the ground state is indeed planar and magnetic,
but the presence of other local energy minima can lead to metastable solutions
with different structural and electronic properties.

\section{Computational details}
In this work, we have investigated carbon vacancies in a graphene sheet, through
first-principles calculations, based on density functional theory
(DFT)\cite{dft}. The DFT calculations were performed using ultrasoft Vanderbilt
pseudopotentials \cite{vanderbilt}, and a generalized gradient approximation
(GGA) for the exchange-correlation potential \cite{perdew}, as implemented in
the VASP code \cite{kresse1,kresse2,kresse3}. In order to study the vacancies we
have used 6x6 supercell with 72 carbon atoms. The vacancies were created by
removing one carbon atom in the graphene sheet. The lattice parameter for
graphene obtained from structural optimization was 2.46 \AA. We have used a
plane-wave-cutoff energy of 400 eV and a Monkhorst-Pack \cite{monkhorst} scheme
with a 5x5x1 k-mesh for the Brillouin zone integration. Calculations were
performed both with and without spin polarization for the graphene sheet
containing a single vacancy. In all calculations the lattice parameter was kept
fixed at the calculated value, whereas the atoms were allowed to relax until the
atomic forces were smaller than 0.025 eV/\AA. 

\section{Results and Discussions}
\begin{figure}[h]
\includegraphics[width= 8.5cm]{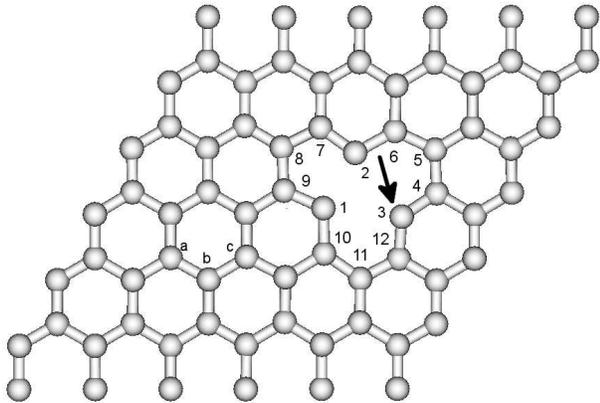}
\caption{The relaxed atomic structure of the graphene sheet with a single
vacancy, where spheres represent the carbons atoms. The arrow indicates the atom
3 with the dangling $\sigma$ bond and atoms 1 and 2 are those rebonded.}
\label{structure}
\end{figure}

Fig.\ref{structure} shows the fully relaxed atomic structure of the graphene
sheet with a single carbon vacancy, obtained by spin-resolved DFTcalculations. 

This structure clearly shows the local breaking down of the threefold symmetry
due to the Jahn-Teller distortion induced by the reconstruction of two
dangling bonds left after removing one carbon atom from the graphene sheet. It
should be noted that, as a result of this distortion, the atoms 1 and 2 move
closer to one another, forming a reconstructed bond with length of 2.07 \AA (as
compared to 2.46 \AA~ in defect-free graphene). This gives rise to the in-plane
displacements of other carbon atoms in the relaxed structure, as shown in Table
\ref{interatomic}.

\begin{table}[h]
\begin{center}
\caption{Distances corresponding to first and second neighbors in a
graphene sheet with a single vacancy and in defect-free graphene, where
$d_{i-j}$ denotes the distances corresponding to different pairs of
atoms as shown in Fig.\ref{structure}.}
\label{interatomic}

\begin{tabular}{cccccc}
\hline\hline
 &        &Graphene-vacancy (\AA)               &        & Graphene (\AA) & \\
\hline
&First-neighbors                            &                \\  
&   &$d_{3-4}$=$d_{3-12}$           =1.37   & &$d_0$=1.42    \\
&   &$d_{2-7}$=$d_{1-9}$            =1.40   & &              \\
&   &$d_{8-7}$=$d_{8-9}$            =1.41   & &              \\
&   &$d_{5-4}$=$d_{11-12}$=$d_{a-b}$=1.42   & &              \\
&   &$d_{6-5}$=$d_{10-11}$=$d_{a-c}$=1.43   & &              \\
&   &$d_{2-6}$=$d_{1-9}$=$d_{10-1}$ =1.47   &&		     \\

\hline
&Second-neighbors                                & \\
&         &$d_{1-2}$=2.07                        & &$d_0$=2.46       \\
&         &$d_{7-9}$=2.37                        & &       \\
&         &$d_{2-8}$=$d_{8-1}$=2.39              & &       \\
&         &$d_{5-3}$=$d_{11-3}$=2.43             & &       \\
&         &$d_{6-4}$=$d_{10-12}$=2.44            & &       \\
&         &$d_{1-3}$=$d_{2-3}$=2.56              & &       \\
&         &$d_{2-5}$=2.58                        & &       \\
&         &$d_{11-1}$=2.59                       & &       \\
\hline
\end{tabular}
\end{center}
\end{table}

Table \ref{interatomic} shows that there exists a variation in the interatomic
distances corresponding to second neighbors, ranging from 2.07 \AA~ up to 2.59
\AA~. The largest change in the in-plane interatomic distances is observed for
atoms 1 and 2 ($d_{1-2}$=2.07 \AA) and is associated with the reconstructed bond
mentioned above. We can also note that the interatomic distance between first
neighbors varies from 1.37 \AA~ up to 1.47 \AA~, compared to 1.42 \AA~ in
defect-free graphene. These results are in good agreement with previous
work \cite{dai,palacios}.

The formation energy of a single vacancy in the graphene sheet at the ground
state of the system was calculated as 
       
 \begin{equation}
 \label{formation}
 E_f=\frac{1}{n}(E_v-\frac{N-n}{N}E_{g}),
\end{equation}

where $N$ denotes the number of C atoms in the defect-free graphene sheet, $n$
represents the number of vacant C atoms (i.e., $n=1$ in the present case)  and 
$E_{g}$ and  $E_v$ are the total energies  of the  defect-free and the
vacancy-containing  graphene sheets, respectively. The formation energy of the
graphene sheet containing a single vacancy calculated according to
eq.(\ref{formation}) was found to be 7.6 eV, which is in good agreement with
the experimental value of 7.0 eV \cite{mayer} and also with previous results of
DFT calculations \cite{dai,faccio2}.

\begin{figure}[h]
\includegraphics[width= 8.5cm]{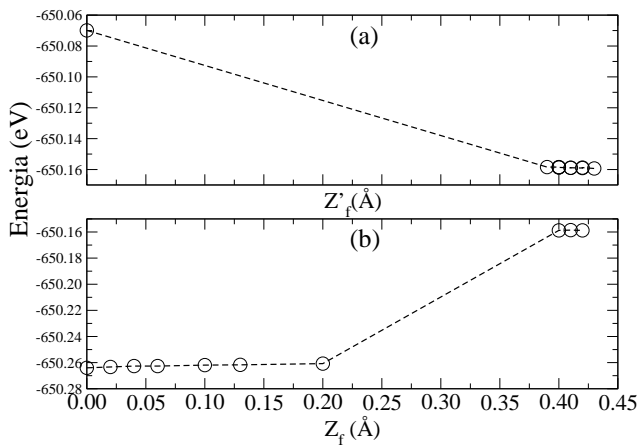}
\caption{Total energy of the graphene sheet containing a single
vacancy calculated as a function of the displacement of atom 3 perpendicular to
the sheet (a) without spin polarization and (b) with spin polarization. The
lines connecting between the points are guides for the eyes.}
\label{shiftz}
\end{figure}

In order to verify the existence of possible metastable solutions involving
different geometries of the graphene sheet containing a single vacancy, we have
investigated in detail the possible displacement of atom 3 (which is the atom
that does not form a reconstructed bond with the other ones, indicated by the
arrow in Fig. \ref{structure}) perpendicularly to the sheet. Then, a series
of calculations were performed, starting from a structure with atom 3 initially
displaced from the sheet by variable distances (named Z$_i$). These calculations
were done for both unpolarized and spin-polarized schemes, with the structure
being allowed to fully relax. The initial (Z$_i$) and final positions (named
Z'$_{f}$ and Z$_{f}$ for the unpolarized and spin-polarized calculations,
respectively) of atom 3 with the corresponding magnetic moments are summarized
in Table \ref{outplane}, whereas the plots of the total energy of the system
as a function of the final position reached by atom 3 are shown in Fig.
\ref{shiftz}.

\begin{table}
\begin{center}
\caption{Final (Z'$_{f}$ and Z$_{f}$ ) distances of atom
3 measured from the sheet after full relaxation for unpolarized and
spin-polarized calculations, respectively. (Z$_{i}$) represents the
initial displacement of atom 3 from the graphene sheet and ($\mu$) is
the magnetic moment corresponding to the relaxed structure.}\label{outplane}
\begin{tabular}{c  c  c c}
\hline\hline
Z$_{i}$ (\AA)   &Z'$_f$	&    Z$_f$ (\AA)     	&     $\mu$ ($\mu_B$) \\
\hline
0.00 		&0.00	&0.00	     	     	& 1.21 \\
0.06 		&0.40	&0.02		     	& 1.21 \\
0.08 		&0.40	&0.05		     	& 1.20 \\
0.10 		&0.41	&0.06	             	& 1.19 \\
0.20 		&0.41	&0.11		     	& 1.16 \\
0.30 		&0.41	&0.13		     	& 1.12 \\
0.40 		&0.41	&0.20		     	& 1.04 \\
0.50		&0.42	&0.41	             	& 0.00 \\
0.60 		&0.42	&0.41	             	& 0.00 \\
0.70 		&0.43	&0.42	             	& 0.00 \\
\hline
\end{tabular}
\end{center}
\end{table}

From the data shown in Table \ref{outplane} we can analyze the behavior of the
final position reached by atom 3 as a function of its initial position. In the
unpolarized case, we can observe that if an initial out-of-plane shift is given
to atom 3, then its final position reaches values around 0.42 \AA (even for
initial values as small as 0.06 \AA). On the other hand, for spin-resolved
calculations there exist different metastable solutions that can be reached
depending on the initial shift. In particular, it should be noted that for
 Z$_i < 0.40$ \AA~ the final position of atom 3 Z$_f$ varies from 0.0 up to
 0.20 \AA~, whereas for Z$_i > 0.40$ the final position is around 0.42
\AA, matching the value found in the unpolarized calculations .

Fig.~\ref{shiftz} shows how the total energy of each relaxed 
structure changes as a function of Z'$_{f}$ or Z$_f$, for calculations
performed without (Figure \ref{shiftz}a) or with (Figure \ref{shiftz}b) spin
polarization, respectively. These results show that the ground state of the
relaxed structure is indeed planar (i.e., Z$_f$ = 0) and spin-polarized, with a
total energy nearly 0.10 eV below that of the unpolarized structure, in good
agreement with some previous results \cite{ma,ugeda2}.

As was already mentioned before, the details of the relaxed structure of
a graphene sheet containing a single vacancy reported in the literature vary
widely. While some authors have reported a planar geometry for the ground state
\cite{nanda,faccio2,ugeda2}, there are also many reports of non-planar
structures with out-of-plane displacements varying from ~ 0.18 to 0.47 \AA~
\cite{barbary,ma,dai}. It is worth emphasizing that in the case of the
unpolarized calculations  (Figure \ref{shiftz}a), atom 3 tends to be shifted by
~ 0.42 \AA~ out-of-plane, whereas the spin-polarized calculations
(Figure \ref{shiftz}b) yield a planar structure. However, it is interesting to
note that even in the case of spin-polarized calculations there exists a
clustering of points close to this value of Z$_f$, suggesting a local
energy minimum in this region. This means that if the initial shift of
atom 3 is above ca. 0.20 \AA~, the structural relaxation leads to metastable
states with Z$_f$ around 0.42 \AA~ and with the same total energy as compared to
the unpolarized structure (ca. 0.10 eV above the ground state), which can help
to understand the reports of shifts with this magnitude found in the literature
\cite{barbary, ma}. Further, it can be observed in Figure \ref{shiftz}b
that a little energy variation and small differences between Z$_i$ and Z$_f$ are
detected among the several final structures obtained for Z$_i$ in the range from
0.00 up to 0.20 \AA~, which is also consistent with other reports \cite{ma,
dai}. These results show that the final reached structures are strongly
dependent on the use of spin polarization in the calculations \cite{faccio2,
ugeda2} and that, even when using spin polarization, the setting of the
initial conditions can significantly change the properties of the relaxed
structure. It is likely that most of the discrepancies found in the literature
regarding the out-of-plane displacements of atom 3 can be associated with the
trends illustrated in Fig.~\ref{shiftz}.

\begin{figure}[h]
\includegraphics[width= 8.5cm]{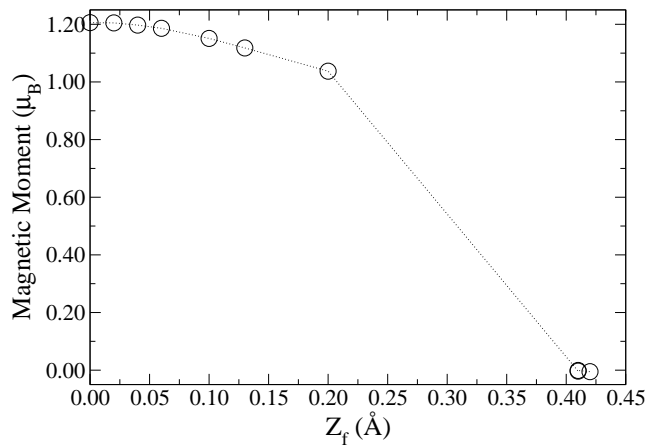}
\caption{Magnetic moment of the graphene sheet containing a single vacancy
calculated as a function of the displacement of atom 3 perpendicular to the
sheet for the relaxed structure. The line connecting the points is a guide for
the eyes.}
\label{magmom}
 \end{figure}

Moreover, the abrupt change of the energy for Z$_f$ $>$ 0.20 \AA~ can be
associated with the reported disconnection of the $\pi$ electron associated with
the vacancy from the 2D electron network \cite{wardana}. For such large shifts,
it is energetically more favorable for the system to accommodate two of the
electrons released by the vacancy in a singlet pair at atom 3, with the other
two electrons becoming part of the reconstructed $\sigma$-like bond between
atoms 1 and 2. Thus, this metastable structure must remain nonmagnetic, with the
same energy (-650.16 eV) irrespective of the use of spin polarization in the
calculations. This behavior can be clearly observed in Figure \ref{magmom},
where the magnetic moment of the graphene sheet containing a single vacancy is
plotted as a function of Z$_f$. It is visible that the planar ground state(Z$_f$
= 0.00 \AA) is magnetic, with a magnetic moment of 1.21 $\mu_B$;
on the other hand, for Z$_f$ $ >$ 0.20 \AA~ the system evolves to a
nonmagnetic state, achieved with Z$_f$ close to 0.42 \AA.

\begin{figure}[h!]
\includegraphics[width= 8.5cm]{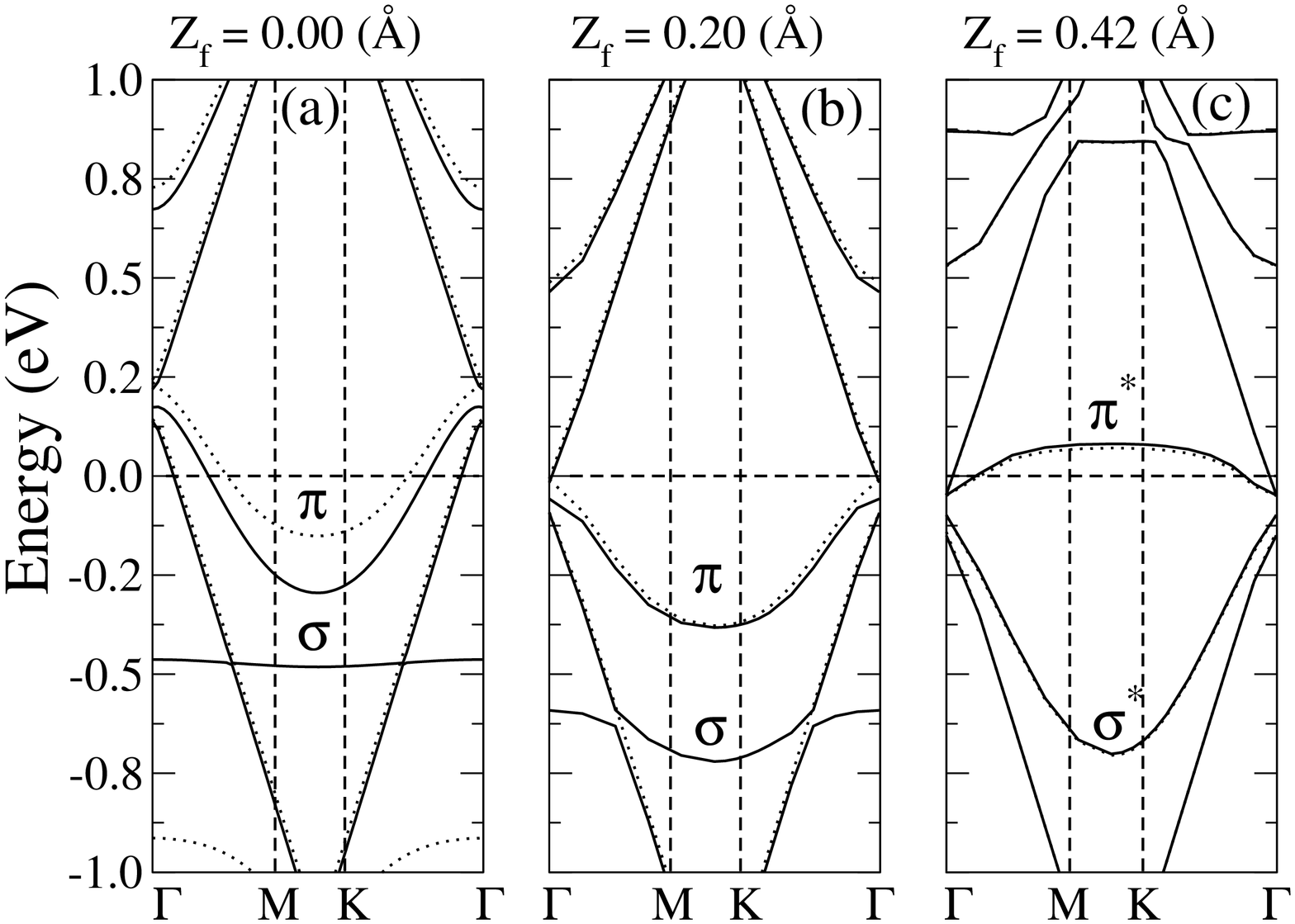}
\caption{Spin-resolved band structures calculated after full relaxation of the
graphene sheet containing a single vacancy for three different values of the
final displacement of atom 3 perpendicular to the sheet: (a) Z$_f$ = 0.00 \AA~;
(b)Z$_f$ = 0.20 \AA~ and (c) Z$_f$ = 0.42 \AA. The zero of energy is at the
Fermi level. Solid and dotted lines represent spin-majority and spin-minority
bands, respectively.}
\label{bands}
\end{figure}

The changes in the electronic structure related to different atomic
configurations as atom 3 is moved away from the graphene sheet are shown in Fig.
\ref{bands}, where we can see the spin resolved band structure for the different
Z$_f$ values, after full relaxation of the system. 

The existence of separated spin-minority and spin-majority bands
is evident in Fig. \ref{bands}a and, to a lesser extent, in Fig. \ref{bands}b,
both for $\sigma$ and $\pi$ bands, corresponding to Z$_f$ = 0.0
\AA~ and 0.20 \AA, respectively. With increasing Z$_f$, one can also observe a
broadening of the $\sigma$ band that was located close to -0.5 eV for
Z$_f$ = 0 and that was associated with a strongly localized state around atom 3
in the planar structure \cite{helm,ugeda2}. At the same time, the spin-minority
$\pi$ band is pushed down and both $\pi$ bands are almost completely below the
Fermi level for Z$_f$ = 0.20 \AA, which leads to a reduction in the magnetic
moment compared to the planar geometry, as shown in Figure \ref{magmom}.
Moreover, Fig. \ref{bands}c shows that for Z$_f$ = 0.42 \AA there exists a
complete overlap of spin-majority and spin-minority bands, leading to the
quenching of the magnetic moment of the sheet. Then, one can note the $\sigma$
and $\pi$ states obtained for Z$_f <$0.20A~ evolved to $\sigma^*$ and
$\pi^*$ states, which are combinations of bonding and antibonding orbitals. This
leads to a stable electronic configuration with double occupancy of the
$\sigma^*$ state and, partially, of the $\pi^*$ states, which are now pinned to
the Fermi level. The quenching of the magnetic moment of the sheet associated
with the out-of-plane displacement of atom 3 has already been previously
reported \cite{ugeda2}.
However, it is not clear whether or not a full relaxation of both atomic and
electronic coordinates was allowed in the structures reported in ref.
\cite{ugeda2}, so it is difficult to quantitatively compare those results with
the ones presented here. But the qualitative mechanism describing the quenching
of the magnetic moment is clearly consistent with the trends shown in Figs.
\ref{shiftz}, \ref{magmom} and \ref{bands}: as atom 3 is shifted away from the
graphene plane, there exists a progressive change of hybridization of that atom
and a mixture of $\sigma$ and $\pi$ states; this changes the occupation of the
corresponding spin-majority and spin-minority bands and thus leads to a partial
compensation of the spin populations and a reduction in the magnetic moment of
the sheet. In the present case, when the displacement of atom 3 exceeds a
threshold close to Z$_f$ = 0.20 \AA~, then a complete compensation of spin
populations in the now mixed states occurs, leading to the settling of a
nonmagnetic state.

\section{conclusion}

DFT calculations have been employed in a systematic investigation of the
effects of out-of-plane atomic displacements on the structural, electronic and
magnetic features of a graphene sheet containing a single vacancy. The results
showed the occurrence of a local distortion around the vacancy in the relaxed
structure, with reconstructed atomic bonds between two atoms close to the
vacancy and with the third atom located in the graphene plane. Total energies,
magnetic moments and electronic configurations of the final structures reached
after full relaxation starting from different initial shifts of this third atom
in the direction perpendicular to the graphene sheet were calculated, both
for polarized and also for unpolarized structures. The present results
showed that the ground state of this system is indeed planar and magnetic, with
spin-resolved $\sigma$ and $\pi$ bands contributing to the total magnetic moment
of the graphene sheet containing a single vacancy. However, we have found the
presence of other local energy minima, which can lead to metastable solutions
with different structural, electronic and magnetic properties that are strongly
dependent on the extent of the out-of-plane shift.

\section*{acknowledgments}

The authors are grateful to Dr. Ricardo Faccio for helpful discussions and
detailed explanations on his own work in the field. The support from Brazilian
agencies CNPq, CAPES and FAPES is gratefully acknowledged.

\section*{references}

\end{document}